# Towards Development of Correct Software using Views

Barbara Paech, Bernhard Rumpe

Institut für Informatik, Technische Universität München

Munich, Germany

http://www.forsoft.de/~(paech|rumpe)

**Abstract**

This paper aims at integrating *heterogeneous documents* used in pragmatic software develpoment methods to describe *views* with a *formal* refinement based software development process. Therefore we propose an integrated semantics of heterogeneous documents based on a common *system model* and a set of *syntactic* development steps with a well-defined semantics for document evolution. The use of the development steps is demonstrated in a small example.

## 1 Introduction

Software engineering methodologies structure the software development into a specific *process* yielding a certain set of *products*. Very often the process is only roughly described (e.g. distinguishing analysis, system design, object design and implementation in OMT[21]), but an elaborate set of description techniques for the products is given (cf. the newly developing standard *Unified Modeling Language* [3]). These description techniques define different *views* on the application system or the software system. While the particular notations and views have changed over the years adhering to a structured, data-oriented or object-oriented paradigma (for a comparison see e.g. [7]), the use of views in software development is indespensable in order to concentrate on different system *aspects* (e.g. static structure, data, behaviour) on different *abstraction levels*. The former reflect concerns of the different participants (e.g. manager, user, system analyst, software designer) and the latter the amount of information relevant in the different stages of software development (e.g. [16]).

Thus the products of *pragmatic* software development methods constitute a quite *heterogeneous* set of documents. *Formal* software development methods, on the contrary, usually offer a *uniform* specification language equipped with constructs for *specification in the large* and a notion of *refinement* (e.g. VDM[2], Z[27] and a great variety of algebraic approaches[26]). In recent years efforts have been made to combine pragmatic and formal methods. However, they have concentrated on giving a formal semantics to pragmatic notations (e.g. [10]), on using pragmatic notations for requirements engineering yielding a formal requirements specification (e.g. [20], [11],[14], [17]) or using the formal specification as an additional means of analysis, design, specification and validation within the pragmatic methods (e.g. [19]).

To our knowledge no proposal has been made for *using heterogeneous documents within a refinement based formal development process*. The present paper is one step in this direction. It advocates

- a *mathematical system model* as a basis for the semantic integration of the heterogeneous documents and

- a set of *syntactic development steps* with a well-defined semantics for document evolution.

In the following we will explain our approach within the context of a software engineering method for distributed object-oriented systems. Thus the system model introduced in section 2 formalizes a system as a set of *interacting components*. In section 3 we define the syntax and semantics of three exemplary view documents: an *object model* for describing the static system structure, *class descriptions* for describing the component interface and *automata* for describing component behaviour. Then we define the syntactic developments steps particular to the individual kinds of documents are defined. In section 5 we sketch over a chain of development steps for the development of a distributed car rental system in order to give a flavour of the handling of our approach. We close with some remarks on related and future work.

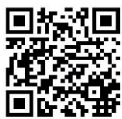

## 2 System Model

The system model serves as a common reference model for the definition of the *semantics* of the description techniques. It therefore is a solid basis for their *integration*. It also helps the different system developers participating in the development process to achieve a common understanding of the kind of system to be developed.

In this section we sketch the system model and motivate its design decisions. The underlying formalism is pure mathematics enhanced with the theory of streams as given in FOCUS ([5]). This allows us to use very powerful and flexible concepts, but all proofs of correctnes over the development steps are paper-and-pencil-proofs. To ensure automatic proof support, the system model, the used description techniques, their semantics and transformations may be coded within a logic, e.g. HOL ([8]) and then proved with a theorem-prover like *Isabelle* ([18]).

A more detailed explanation of a more general system model, allowing combined use of hardware and software components can be found in [24] and [12].

*Outline of the System Model*

The system model aims at describing distributed object-oriented systems. It formalizes a system as a set of *interacting components*, called *objects*. All objects interact by *asynchronous message passing*. Data (states) of objects are encapsulated, no sharing occurs. The concept for message addressing uses identifiers of objects. Dynamic creation of components is modeled by using possibly infinite sets of objects, which are "activated" by special messages.

We do not abstract from time, because this on the one hand allows us to describe real-time object-oriented systems and on the other hand prevents semantic problems as e.g. the merge anomaly ([4]).

*Formalisation of the system model*

The system model $\mathbb{SM}$ formally is the set of systems, we are interested in. A system $sys \in \mathbb{SM}$ is a tuple consisting of all the sets and functions defined in the following.

Let $VAL$ denote the universe of all *values* in system $sys$. We do not regard objects as values, but their state is composed of values, and an object identifier is a value as well. This universe is structured by the set $SORT$ of sorts, where each sort $s \in SORT$ denotes a subset of $VAL$ by the function $values$:

$$values(s) \subseteq VAL.$$

Let $VAR$ be the set of variable names including attributes. We assume each variable name to be used only once, and therefore allow an assignment of a sort for every variable, denoted by:

$$sort : VAR \rightarrow SORT.$$

*Object and Class Signature*

An object-oriented system consists of a set of objects, that are conceptually and/or spatially distributed. Every object is uniquely identified by an element of the enumerable set $ID$ of *identifiers*.

Objects with common behavior and interface are grouped together by the notion of *class*. A finite set of classes thus defines a type system for objects[1]. Each system contains a set of class names $CN$. One of the purposes of a class is, that it serves as a sort of its object identifiers. This is very similar to the class notion in $C++$, where the object type `OBJECT` and the type of object identifiers `OBJECT &` are both used. We therefore set

$$CN \subseteq SORT$$

---
[1] For sake of brevity we have omitted inheritance. An example of a system model with inheritance can be found in [22].

and demand that each class-sort $c \in CN$ contains only identifiers as its values:

$$\forall c \in CN. values(c) \subseteq ID.$$

Vice versa each object identifier $id \in ID$ has a unique class assigned via the function *class*. Object $id$ is *instance* of class $class(id) \in CN$.

The *signature* of an object is given by the set of messages it can accept and emit. Objects which are instances of the same class have a common signature defined by a finite set of method names, the arity and the kind of their parameters. We do not define the concrete signature of a class here. But let us assume $\Sigma_c$ denotes the appropriate set of input messages of class $c$.

Messages in the system model contain the identifier of the receiving object. As this is not part of the message contents, we define the set of input messages of object $id \in ID$ as

$$In_{id} = \{(id, msg) | msg \in \Sigma_{class(id)}\},$$

where the message body $msg$ may be itself a complex structure composed of a method name and arguments. As the actual set of output messages of one object is determined by the implementation respectively the specification itself, we use the set $MSG = \bigcup_{id \in ID} In_{id}$ of all messages as output interface.

*Black-Box Behaviour*

Let $M^{\overline{\omega}}$ denote an infinite timed stream over $M$. The *black-box behavior of object* $id$ is described as a function $behavior_{id}$ relating a stream of input messages to a stream of output messages[2]

$$behavior_{id} \subseteq In_{id}^{\overline{\omega}} \xrightarrow{p} \wp(MSG^{\overline{\omega}}).$$

Similarly to message signatures, each class is given an attribute signature denoted by $\Theta_c \subseteq VAR$, which assigns a set of variables to each class.

*Object Communication*

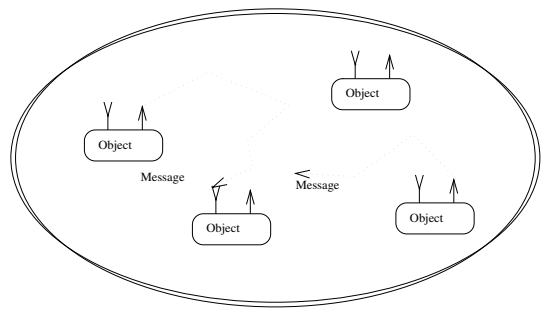

Figure 1: Object communication

Every kind of computation is performed within objects. However these objects are not directly composed. As shown in Figure 1, this is done by embedding all objects in a *communication medium* that does the necessary routing of messages. As communication is asynchronous in general, the necessary buffering of messages is done within the communication medium as well. The communication medium has to obey several restrictions, e.g. messages are not lost or duplicated and order has to be maintained between two messages with common source and destination. In principle the communication medium is allowed to delay messages, but not infinitely long. If one is interested

---

[2] $\xrightarrow{p}$ denotes a *pulse-driven* function ensuring that at any point in time the input does not depend on the future

in specifying real-time object-oriented systems, the medium may be refined to a medium without as well as with a restricted sort of delay. Although the medium is not a component that is intended to be implemented, we can describe the communication medium as an ordinary component, as done in [12].

*State-Box Behavior*

Besides the object signature that constitutes the interface, objects also have an internal state, composed of a finite set of attributes. The behaviour of an object can be given either in a property oriented way as a black-box or based on internal states.

The state space of an object is given by

$$states_{id} \subseteq (\Theta_{class(id)} \to VAL) \times CTRL.$$

The value part is determined by its attributes. The control part $CTRL$ resembles an abstraction of the program counter and is not further determined here. A set $states_{id}^0$ contains the initial states of the component $id$:

$$states_{id}^0 \subseteq states_{id}.$$

To describe the behavior of a component $id$, a state transition relation $\delta_{id}$ is used. Each transition consists of a source state, a sequence of accepted input messages, a sequence of emitted output messages and a destination state. The transition relation is nondeterministic allowing also for underspecification of the object:

$$\delta_{id} : states_{id} \times In_{id}^* \times MSG^* \times States_c.$$

$(In_{id}, MSG, states_{id}, states_{id}^0, \delta_{id})$ constitutes a timed port automaton (defined in [9]). Timed port automata have a precisely defined black-box semantics and therefore fully determine $behavior_{id}$.

To give a full account of the system behaviour object creation and deletion must be handled. As mentioned before, this is modeled by special messages. The details are omitted here. A *system run* is described by the stream of messages each object accepts and emits, as well as by the sequence of states each objects assumes.

## 3  Documents and their semantics

In this section we demonstrate, how different description techniques, like object models and automata can be given an integrated formal semantics based on the system model. The common semantic function is denoted as

$$[[.]] : \mathcal{DOC} \to \wp(\mathbb{SM})$$

associating with each document a set of system models satisfying the specification given in the document.

For each kind of documents we give

- a concrete (usually graphical) notation,

- an abstract syntax definition and

- a semantics definition.

The concrete, at least in part graphical notation, is used by the software engineer. The abstract syntax comprises this notation without syntactic sugar. It is sufficient to define the semantics based on the later.

## 3.1 Object Models

An object model defines the structure of the system in terms of *classes* and *data relationships*. As graphical notation we use a kind of entity relationship diagram where boxes denote classes and lines data relationships. Lines are labelled with role identifiers and the cardinality ($*$ denotes 1:n, no label denotes 1:1).

Figure 2 shows the initial object model of the example distributed car rental system. This system consists of different *branches* where cars can be picked-up at one branch and returned at a possibly different one. For each rented car the *rental* details are stored. Here cars are not modelled as objects, but as attributes of *branches* and *rentals* (see the class description of figure 3). The initial object model contains the classes *branch* and *rental* and data relationships to record the *pick-up branch* and *return branch* of each *rental*.

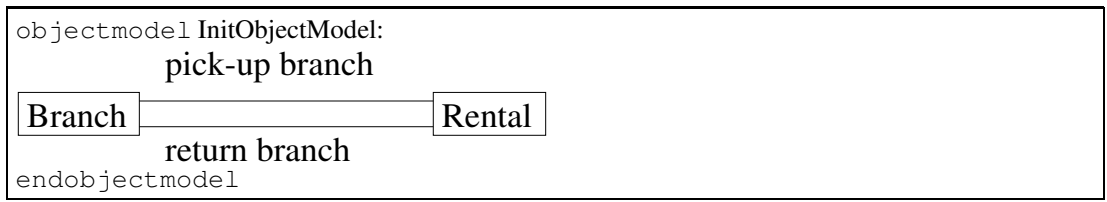

Figure 2: Initial Object Model

The abstract syntax of an *object model* document is given as the tuple $(C, R)$, consisting of

- a set of classes $C \subseteq CN$ and
- a set of data relationships $R \subset Role \times Role$

where a role $(c, rn, card) \in Role$ consists of a classname $c \in C$, an optional rolename $rn$ and a cardinality $card \in \{1, *\}$.

The semantics $[[(C, R)]]$ of object model $(C, R)$ is defined as the set of systems $sys \in \mathbb{SM}$ which fulfill the following properties:

1. Classes exist: $C \subseteq CN$.

2. The semantics of a data relationship is given by using rolenames as attributes to refer to an object in case of cardinality 1 and to a set of objects in case of cardinality $*$. The full semantics is more involved, since cardinality restricts the set of possible system states.

## 3.2 Class Descriptions

A class description defines the signature of classes in terms of methods and attributes.

Figure 3 shows the class descriptions for *branch* and *rental*. They are based on some type definitions which are collected in a special document. These data type definitions constrain the sorts $SORT$ of the system model. Their syntax and semantics is straightforward and not included here for sake of brevity. In the example, methods are not used in these initial class descriptions. A class description using methods is given in figure 5 in section 5.

Thus the abstract syntax of a *class description* is given as a tuple $(c, meth, attr)$, consisting of

- a classname $c$,
- a set of method signatures $meth$ and
- a set of attributes $attr$.

```
classdocument InitBranch :
   class Branch ;
         attributes town : Town ;
                    available_cars : Set Car;
endclassdocument
```

```
classdocument InitRental :
   class Rental ;
         attributes begin : Date ;
                    end : Date ;
                    status : Rental_Status ;
                    car : Car;
                    pick-up_branch : Branch ;
                    return_Branch : Branch;
endclassdocument
```

Figure 3: Initial class descriptions

The semantics $[[(c, meth, attr)]]$ of class description $(c, meth, attr)$ is given accordingly as the set of systems $sys$ where:

1. The class exists: $\{c\} \in CN$.

2. Method signatures are defined[3]: $Messages(meth) \subseteq \Sigma_c$.

3. Attribute signatures are defined: $attr \subseteq \Theta_c$.

The main decision is the way we cope with *absent information*. The semantics of a document is given in a *loose style*. The absence of a piece of information does not imply that it *must* be absent. This allows to later add further details without changing but only detailing the semantics of a document. For example, further classes and relationships can be added, because the semantics does not rule out their existence. This is achived by using subset relations rather than equalities in the semantics definitions and is the basis for a powerful *refinement calculus* comprising the development steps, a software engineer uses.

## 3.3 Automata

Automata are a well suited concept to give a state-based description of object behavior. However, a lot different automata variants are used for different purposes (e.g. I/O-Automata ([15])). We use automata to describe the lifecycle of objects. The lifecycle determines the sequence of states an object passes through, the sequence of inputs it accepts and the sequence of output it emits. An automaton describes object states as nodes and the processing of each input message by a transition. One should note that in contrast to the use of automata in established object-oriented description techniques like OMT[21] we incorporate result messages of object calls as methods into the lifecycle. This is necessary to give a formal semantics to the description technique.

---

[3]Given a concrete formal notion of the syntactical interface $meth$, the function $Messages$ determines a minimal set of messages for $\Sigma_c$.

The set of possible object states and transitions is usually infinite. We therefore use an appropriate *abstraction mechanism* to allow a finite representation of an automaton. This abstraction is done by grouping *object states* into equivalence classes, called *automata states*. Accordingly transitions are grouped.

To define abstractions and to keep the possibility of defining fine grained behavior, a logic or an algebraic specification language, for example SPECTRUM ([6]) or HOL ([8]) is used. This allows to define state predicates to determine equivalence classes of states and to define transitions by pre- and postconditions. Let us assume, a logic language $\mathcal{L}$ is given. We write $\mathcal{I}_{sys}^{\alpha}(f)$, if formula $f \in \mathcal{L}$ is valid in system $sys$ under variable assignment $\alpha : VAR \to VAL$.

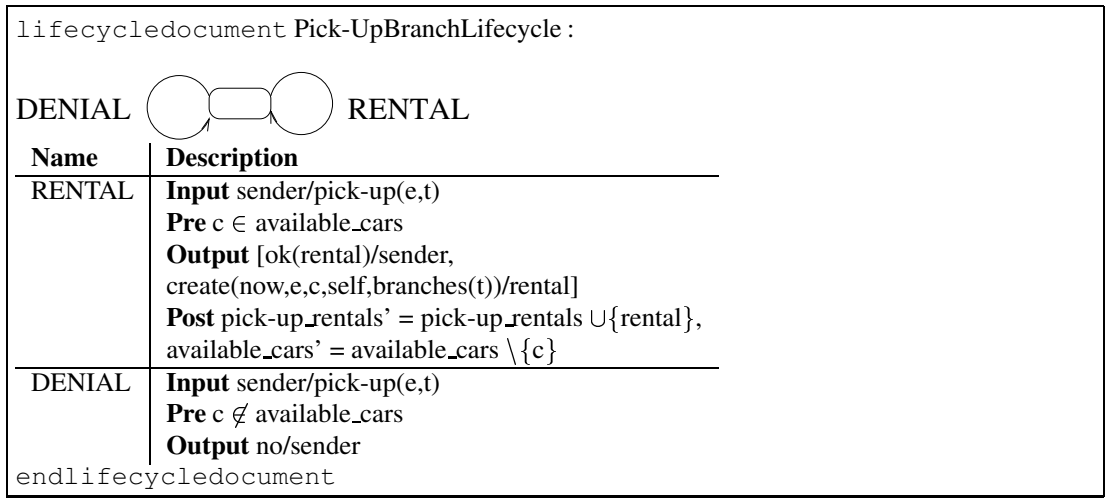

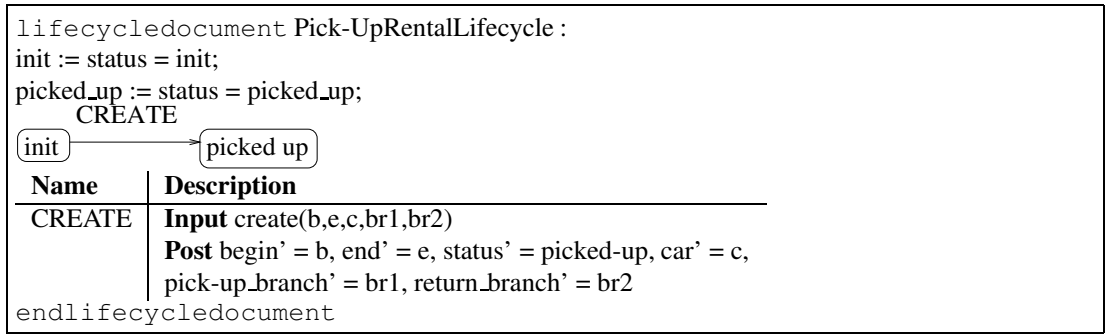

Figure 4: Pick-up Lifecycles

Figure 4 shows the lifecycles of *branch* and *rental* representing the interaction necessary for the *pick-up* functionality of the system.

In the graphical representation each transition is given a name and labelled with an input message, a sequence of output messages, a precondition and a postcondition. Input and output messages may have free variables, whose values are determined from the conditions. We use the notation $sender/message$ and $message/receiver$ to indicated the sender and receiver of messages. The precondition also restricts the source state. The postcondition restricts the destination state, denoted as primed attributes as in TLA[13]. States are labelled with predicates defined in terms of attributes of the corresponding class.

In the example, upon receipt of the message *pick-up(e,t)* the *branch* distinguishes two cases: if a *car* is *available*, it creates a *rental*, updates it *pick-up rentals* and the *available cars* and acknowledges the *rental* to the *sender*. Otherwise, it denies the *rental*. The *rental* reacts in the initial lifecycle just to the *create* message by initializing its attributes.

In the abstract syntax, we deal only with pre- and postconditions, because input messages like $m(a,b)$ can be added to the precondition as $\exists a, b. in = m(a,b)$ and similarly for the output. Conditions then only have four free variables: $in$ for the input message, $out$ for the output message sequence, and $s$ resp. $t$ for source and destination state. Preconditions only use $in$ and $s$, state conditions only use $s$. A precise definition can be found in [22].

The abstract syntax of a *lifecycledocument* is given as the *automaton* $(c, S, \Lambda, \delta, s^0)$, consisting of

- a class $c \in CN$ the automaton belongs to,
- a nonempty set of automata states $S$,
- a map $\Lambda : S \to \mathcal{L}$, assigning a state predicate to each automaton state,
- a transition relation $\delta \subseteq S \times \mathcal{L} \times S \times \mathcal{L}$, and
- an initial state $s^0 \in S$.

Several context conditions for automata have to be fulfilled, to make automata an intuitive description technique. The most interesting ones are:

1. Each automaton state $s$ coresponds to a nonempty set of object states, i.e.
$$\exists \alpha. \mathcal{I}^\alpha_{sys} \Lambda(s).$$

2. State predicates are disjoint and thus denote equivalence classes of object states, i.e.
$$(\exists \alpha. \mathcal{I}^\alpha_{sys} \Lambda(s) \wedge \mathcal{I}^\alpha_{sys} \Lambda(t)) \Rightarrow s = t.$$

3. Transitions have to be enabled. This means, that the firing of transition $(s, pre, t, post) \in \delta$ is determined by the precondition $pre$, and the postcondition $post$ can always be fulfilled if the precondition is true[4]:
$$\forall \alpha : \{s, in\} \to VAL. \mathcal{I}^\alpha_{sys} pre \wedge \Lambda(s) \Rightarrow \exists \beta : \{s, out\} \to VAL. \mathcal{I}^{\alpha+\beta}_{sys} post \wedge \Lambda(t).$$

## 4 Document Evolution

In this section we introduce the refinement rules for the documents. First, the basic document relations are defined for an arbitrary set $\mathcal{DOC}$ of documents. These relations are based on the common semantic function $[[.]] : \mathcal{DOC} \to \wp(\mathbb{SM})$ associating with each document a set of system models satisfying the specification given in the document. This function is straightforwardly extended to subsets $D \subset \mathcal{DOC}$ such that

$$[[D]] \stackrel{\text{def}}{=} \bigcap_{d \in D} [[d]].$$

Consistency and refinement of documents are defined as usual.

Given a semantics based refinement notion $\models$, we are now interested in establishing a syntax based refinement calculus $\vdash$ that allows to manipulate documents in such a way, that each manipulation is a correct refinement. This means:
$$\forall D, d. D \vdash d \Rightarrow D \models d$$

---

[4] With $\Lambda(t)$ we denote the logical formula, where each occurence of a free variable (here only $s$) is replaced by its primed variant (here $s$)

As we have a heterogeneous set of description techniques, we naturally establish a set of refinement rules ⊢ rather than one. For practical usage it is not necessary to have a complete set of rules. Instead one should give a comprehensive set of basic rules, that can be combined to more powerful ones covering the standard ways of constructing documents incrementally . ⊢ is defined as the smallest transitive relation, that incorporates the set of given basic rules. In the following, we define the rules for the description techniques introduced before.

⊢ is itself a heterogeneous relation. It is important to relate abstract specifications, such as automata to concrete ones, such as code in a executable language. Moreover, ⊢ captures the notion of code generation. For example, if an automaton has only executable preconditions and a restricted form of postconditions, it can be automatically translated to code.

*Object model and Class Description*

Based on the semantics of object models, a refinement calculus can be established. It consists of the following atomic rules:

**addclass(c)**  A new class $c$ may be added.

**addrel(rel)**  A new data relationship may be added.

**refrel(oldrel,newrel)**  A data relationship may be refined, e.g. by restricting a cardinality or adding a rolename.

Class descriptions themselves can be refined by these rules:

**addmeth(m)**  The list of message signatures may be extended by a method $m$.

**addattr(a)**  The list of attributes may be extended.

It can be easily proven, that the defined steps are correct.

*Automata*

For automata, the refinement calculus is more difficult, as proof obligations have to be generated to ensure that the resulting automaton fulfills the context conditions given in section 3.3. These proof obligations have to be proven with a theorem prover in order to ensure the correctness of the rule application.

In [22] and [23] a comprehensive set of refinement steps is given and proven correct. It consists of:

**addstate(s)**  New automaton states may be added, denoting equivalence classes of object states that where previously unreachable.

**remstate(s)**  Unreachable automaton states may be removed.

**split(s)**  Automaton states may be splitted with splitting transitions accordingly.

**addtrans(t)**  Transitions may be added, if the input of the new transitions could not be processed by given transitions before.

**remtrans(t)**  Transitions can be removed (see below).

**reftrans(t)**  Transitions can be refined, e.g. splitted or the postcondition strengthened.

**reminit(s)**  Initial states may be removed.

Proof obligations are omitted here.

Each of these steps makes a behavior description more deterministic or more detailed.

**Towards Development of Correct Software using Views**

# 5 A Development Example

In this section we show by way of example how to employ the refinement rules in the development process.

The example consists of the development process for the distributed car rental system introduced earlier. The process is roughly structured as follows. Starting from an initial object model containing *branches* and *rentals*, initial class descriptions and type descriptions are developed. Then the *pick-up* functionality of the *branch* is developed, followed by the *return* functionality.

The initial object model and class descriptions were shown in figure 2 and 3 respectively.

Now the *pick-up* functionality of branches is developed. Therefore a method *pick-up* with parameters *end* and *return town* is introduced. To allow the *branch* to look up the *branch* resident in the *return town*, it must incorporate a corresponding catalogue *branches*. Also the *rentals* processed at the branch are recorded. The *pick-up* method checks, whether a car is available. If so, it creates a corresponding rental. Otherwise the pick-up request is denied. Figure 5 shows the refined object model and class description for branch. The corresponding *branch* and *rental* lifecycles were shown in figure 4. Also the class description for *rental* is refined by adding the method *create* which initializes the attributes. Both documents are omitted here for sake of brevity.

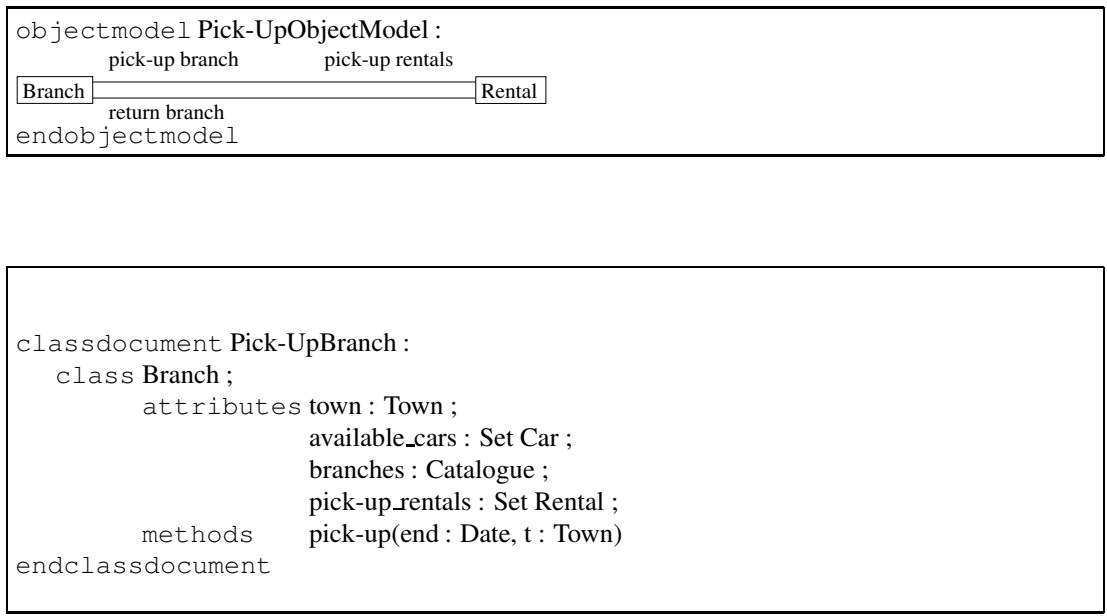

Figure 5: Pick-up object model and class descriptions

It is straightforward to show that the *pick-up* funtionality is a refinement of the initial documents: In the object model a role name is added, therefore *Pick-UpObjectModel* is constructed from *InitObjectModel* by applying *refrel(pick-up branch, pick-up rentals)*. In the class diagram only attributes and a method were added. Therefore *Pick-Up Branch* is constructed from *InitBranch* by applying *addattr(branches), addattr(pick-up rentals), addmeth(pick-up)*. The *rental* documents are constructed similarly.

Following the above development steps the *return* functionality is developed. The object model and the class descriptions are refined by introducing an additional attribute in *branch* to store the *returned rentals* and by introducing a *return* method at the branches. In reaction to the *return* message the *return branch* asks the *rental* for the identity of the pick-up branch. It processes the answer with the method *inform* which sends the *car* back to the *pick-up branch*. The latter is processed by a method *car return* in *branch*. Also *rental* is extended by a method *return* which updates the status of the *rental* and sends the identification of the *pick-up branch* to the *return branch*. In figures 6 and 7 the refined descriptions are given. In the lifecycle documents only the new methods are listed.

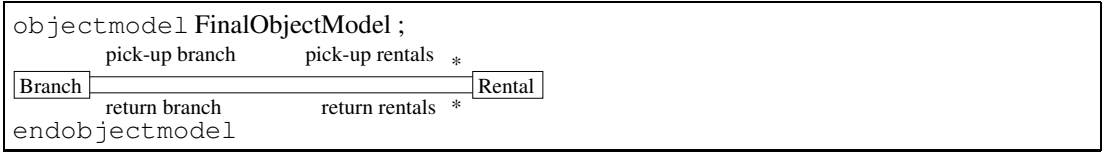

```
objectmodel FinalObjectModel ;
endobjectmodel
```

```
classdocument   FinalBranch :
   class Branch ;
        attributes Town : Town ;
                   available_cars : Set Car ;
                   branches : Catalogue
                   return_rentals : Set Rental ;
                   pick-up_rentals : Set Rental ;
        methods    return(r : Rental, c : Car);
                   pick-up(end : Date, t : Town);
                   inform(pu_branch : Branch, c : Car);
                   car_return(r : Rental, c : Car)
endclassdocument
```

```
classdocument   FinalRental :
   class Rental ;
        attributes begin : Date ;
                   end : Date ;
                   status : Rental_Status ;
                   car : Car;
                   pick-up_branch : Branch ;
                   return_Branch : Branch;
        methods    create(b : Date, e : Date, c : Car, pub : Branch, rb : Branch);
                   return(c : Car)
endclassdocument
```

Figure 6: Final object model and class descriptions

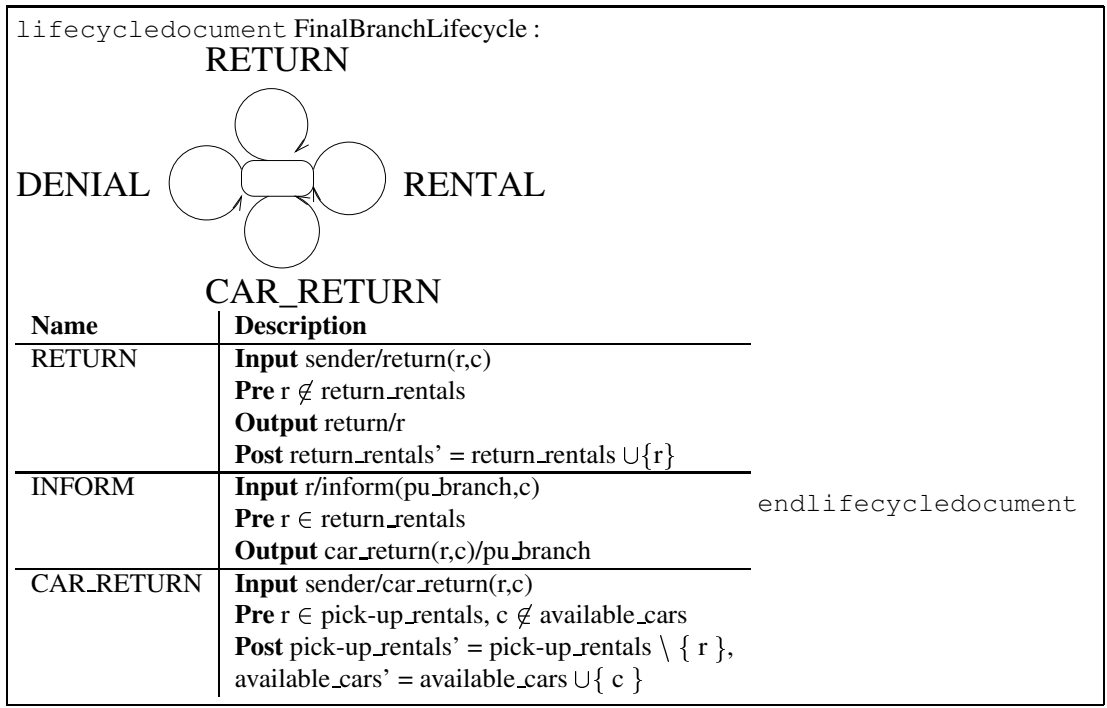

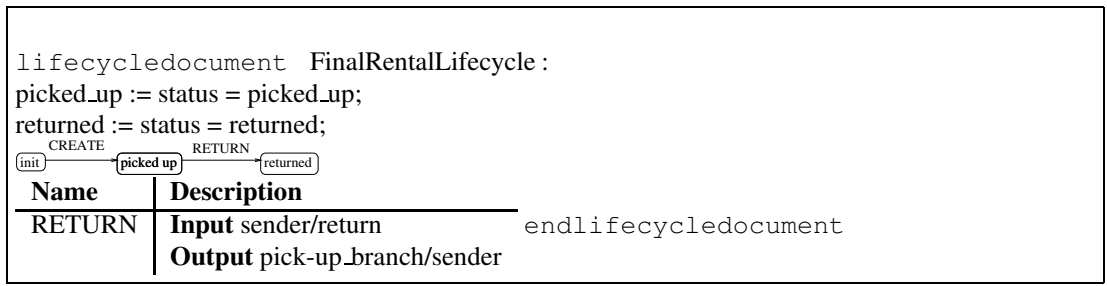

Figure 7: Final lifecycles

It is straightforward to show that the final object models and class descriptions refine the *pick-up* documents by using *refrel, addattr* and *addmeth*.

For the *branch* lifecycles, the involved refinement steps are just the addition of transitions. Since the state label is *true* and the transitions add reaction to new method calls, the context conditions are satisfied.

For the *rental* lifecycles addition of state and transitions is involved. Again, the context conditions are trivially satisfied.

# 6 Conclusions

We have shown a first step to adapt the formal development process by refinement to the pragmatic development process with views. The main elements of our approach are the mathematical system model and the development calculi for the view documents. These features are also necessary to give a powerful tool support to pragmatic methods. Todays CASE-Tools only offer the functionality of graphical editors for the view documents together with a repository of documents to allow import checks. Based on our approach *consistency* checks are possible using the common semantical model. *Correct Development steps* are made practical through guidance by the development calculi.

There is a bulk of research on software process modelling trying to give tool support to the activities of the development process (e.g. [1]). However, in this research the correctness of the development steps is not treated.

In the area of formal software development the most similar approach is KORSO[20]. However, it does not deal with heterogenous documents.

*Future Work*

There remains much to be done to apply this approach to a complete set of view documents as e.g. proposed in UML [3]. On one hand there are some open questions regarding the semantic foundation of views describing object interaction (e.g. operation specification by pre and postconditions, collaboration diagrams), as discussed in our UML formalization [?]. On the other hand it will require some effort to devise the development calculi for these views. Even more effort will be necessary to build a tool supporting the refinement steps together with a full fledged version control system like RCS [25].

# Acknowledgements

For a careful reading of a draft version of the paper we thank Manfred Broy and Cornel Klein. This paper originated in the SYSLAB project, which is supported by the DFG under the Leibnizpreis and by Siemens-Nixdorf.